\documentclass[aps, prl, superscriptaddress, twocolumn, tightenlines, showpacs, floatfix, pdftex]{revtex4-1}
\usepackage[T1]{fontenc}
\usepackage[normalem]{ulem}
\usepackage{textcomp}
\usepackage{mathrsfs,mathtools,xfrac}
\usepackage{amsmath,amssymb,amsthm,amscd,bm}
\usepackage{txfonts} 
\usepackage{afterpackage,multirow,comment}
\usepackage{array,booktabs,minibox}
	\newcolumntype{P}[1]{>{\centering\arraybackslash}p{#1}} 
\usepackage[dvipsnames]{xcolor}
\usepackage{tikz}
\usepackage{float,graphicx,subfigure}
\usepackage[colorlinks,bookmarks=false,citecolor=blue,linkcolor=blue,urlcolor=blue,filecolor=blue]{hyperref}
\usepackage{wasysym}

\usepackage {colortbl} 
\usepackage{lipsum}

\DeclareRobustCommand{\rchi}{{\mathpalette\irchi\relax}}
\newcommand{\irchi}[2]{\raisebox{\depth}{$#1\chi$}} 

\begin{document}
\title{Hybrid Symmetry Breaking in Classical Spin Models With Subsystem Symmetries}

\author{Giovanni Canossa}
\affiliation{Arnold Sommerfeld Center for Theoretical Physics, University of Munich, Theresienstr. 37, 80333 M\"{u}nchen, Germany}
\affiliation{Munich Center for Quantum Science and Technology (MCQST), Schellingstr. 4, 80799 M\"{u}nchen, Germany}

\author{Lode Pollet}
\affiliation{Arnold Sommerfeld Center for Theoretical Physics, University of Munich, Theresienstr. 37, 80333 M\"{u}nchen, Germany}
\affiliation{Munich Center for Quantum Science and Technology (MCQST), Schellingstr. 4, 80799 M\"{u}nchen, Germany}
\affiliation{Wilczek Quantum Center, School of Physics and Astronomy, Shanghai Jiao Tong University, Shanghai 200240, China}

\author{Ke Liu}
\email{ke.liu@lmu.de}
\affiliation{Arnold Sommerfeld Center for Theoretical Physics, University of Munich, Theresienstr. 37, 80333 M\"{u}nchen, Germany}
\affiliation{Munich Center for Quantum Science and Technology (MCQST), Schellingstr. 4, 80799 M\"{u}nchen, Germany}

\date{\today}
\begin{abstract}
We investigate two concrete cases of phase transitions breaking a subsystem symmetry.
The models are two classical compass models featuring line-flip and plane-flip symmetries and correspond to special limits of a Heisenberg-Kitaev Hamiltonian on a cubic lattice.
We show that these models experience a hybrid symmetry breaking by which the system display distinct symmetry broken patterns in different submanifolds.
For instance, the system may look magnetic within a chain or plane but nematic-like when observing from one dimensionality higher.
We fully characterize the symmetry-broken phases by a set of subdimensional order parameters and confirm numerically both cases undergo a non-standard first-order phase transition.
Our results provide new insights into phase transitions involving subsystem symmetries and generalize the notion of conventional spontaneous symmetry breaking.
\end{abstract}
\maketitle

\section{Introduction}
Subsystem symmetries are symmetries acting on lower-dimensional manifolds of the system, such as lines or planes of a cube.
Their early studies in condensed matter physics are in the context of compass models~\cite{Nussinov15} which represent a family of effective Hamiltonians for direction-dependent spin-orbital couplings in correlated electron systems~\cite{Kugel82, Brink04, Khaliullin05} and ultracold quantum gases~\cite{Duan03, Wu08}.
The interest is recently renewed thanks to the developments of fracton topological orders~\cite{Chamon05, Haah11, Vijay16, Nandkishore19, Pretko20} that can be obtained by gauging certain subsystem symmetries.
These symmetries have fundamentally different physical implications than global and gauge symmetries.
A global symmetry has a constant ground-state degeneracy (GSD), and the Landau order-parameter theory of spontaneous symmetry breaking describes its phase transition~\cite{BookAnderson}.
A gauge symmetry supports topological degeneracies on closed manifolds and fractionalized excitations, while its phase transition is typically formulated as topological field theories~\cite{Kogut79, Wen90, Kitaev03}.
In contrast, the GSD arising from subsystem symmetries scales subextensively with the system's size.
Their field theories can mix IR and UV physics~\cite{Seiberg21, Seiberg20, Seiberg21b}, while their ordering mechanisms and phase transitions do not fully fit into any known scenarios~\cite{Lake21,Rayhaun21,Distler22}.

In this work, we study two classical 3D lattice models from an order-parameter perspective to explore the impacts of subsystem symmetries on phase transitions.
The models belong to the compass model family and exhibit $Z_2$ line-flip or plane-flip subsystem symmetries.
They do not support fracton excitations by construction and hence can untwist the influences of subsystem symmetries and fractons.

We find that breaking these subsystem symmetries leads to distinct long-range orders in different submanifolds of the system, which we refer to as \emph{hybrid symmetry breaking}.
The phase transitions in both cases are first order; nevertheless, their finite-size dependences differ from the conventional first-order scaling.
    
The manuscript is organized as follows. In Section~\ref{sec:model_op} we define the models and construct their ground states and order parameters.
Section~\ref{sec:scaling} is devoted to a brief review of the standard and non-standard finite-size scalings for first-order phase transitions.
Section~\ref{sec:anti-Kitaev} and Section~\ref{sec:Kitaev} discuss the phase transitions and the resulting orders breaking the planar and linear subsystem symmetries, respectively.
We conclude in Section~\ref{sec:conclusion} with an outlook.

\begin{figure}[!t]
\centering
\includegraphics[width =.48\textwidth]{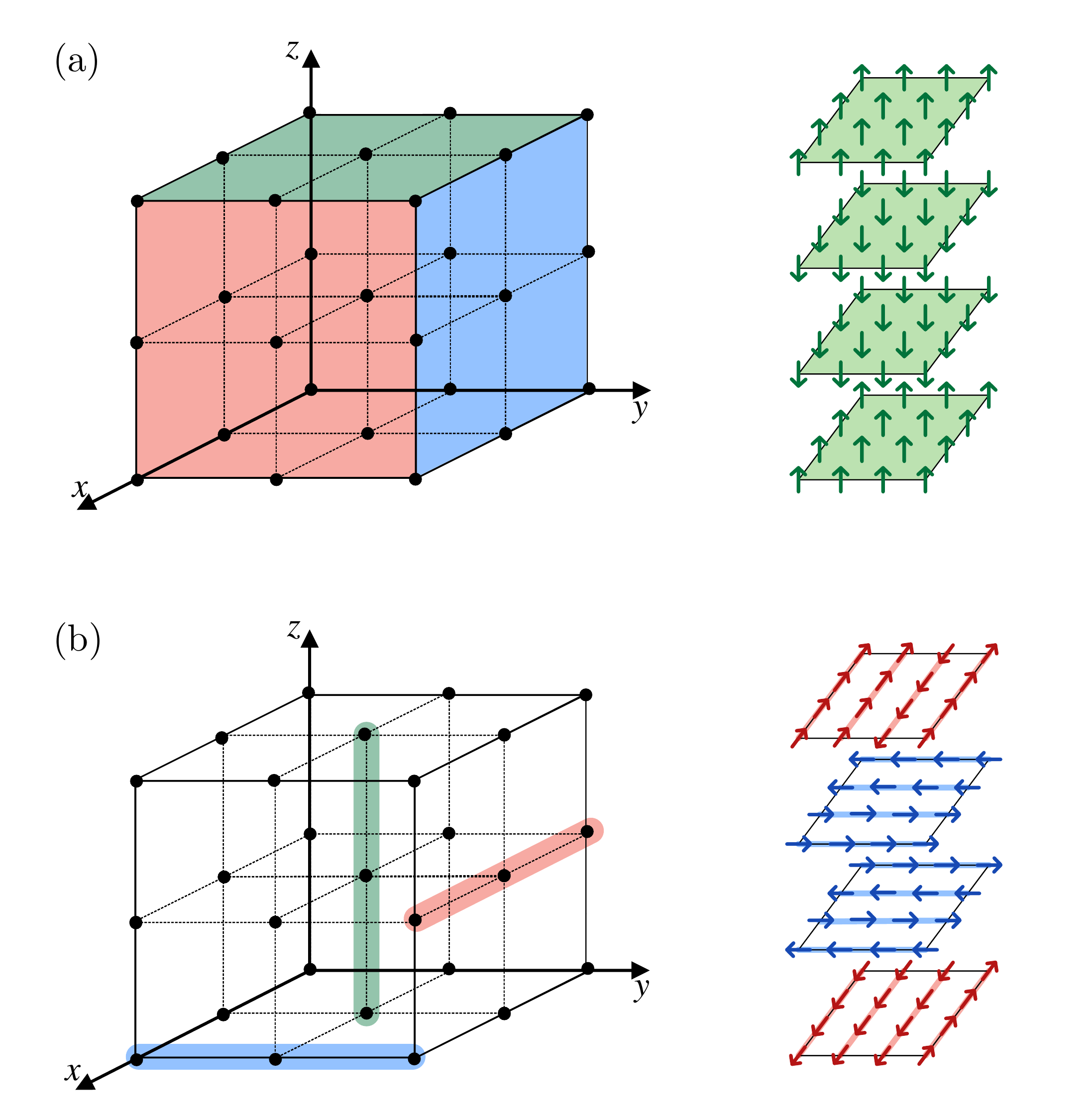}
\caption{Illustration of the models and ground states. (a) In the anti-Kitaev limit, interactions between $S^{\gamma}$ components are canceled on $\gamma$ bonds. The reduced Hamiltonian is invariant under flipping $S^{\gamma}$ for all spins in a spatial plane perpendicular to $\hat{\mathbf{e}}_\gamma$. A ground state is illustrated in the right panel: the system spontaneously selects a global ordering direction $\hat{\mathbf{e}}_\gamma$ and each plane of spins orthogonal to it develops magnetic order along $\pm\hat{\mathbf{e}}_\gamma$.
(b) In the Kitaev limit, spins interact solely through  $S^{\gamma}$ components along the direction $\hat{\mathbf{e}}_\gamma$. The model accordingly gains a line-flip symmetry. Ground states are comprised of magnetically ordered lines arranged into planes with rank-2 nematic order.
Red, green, and blue colors distinguish orientations of lines and planes, and $\hat{\mathbf{e}}_\gamma$ denotes a unit lattice vector.
}
\label{fig:model}
\end{figure}

\section{Models and order parameters}\label{sec:model_op}
One of the simplest models permitting subsystem symmetries is the classical Heisenberg-Kitaev Hamiltonian on a 3D cubic lattice,
\begin{align} \label{eq:model}
H =  \sum_{\langle ij \rangle_\gamma} \left( J \vec{S}_i \cdot \vec{S}_j + K  S_i^{\gamma} S_j^{\gamma} \right),
\end{align}
where $\vec{S}_i$ is a classical $O(3)$ spin at lattice site $i$, $\gamma = x,y,z$ labels its three components,
and $\langle ij \rangle_\gamma$ denotes a nearest-neighboring bond along the lattice direction $\hat{\mathbf{e}}_\gamma$.
Hence, the interaction in the second term is directional.
This Hamiltonian generally has a global cubic symmetry and magnetic ground states.
The linear and planar subsystem symmetries of our interest and supporting non-trivial ground states are present in the limiting cases $J = 0$ and $J+K =0$, respectively.

For simplicity, we consider a ferromagnetic $K=-1$ at the $J=0$ limit and a ferromagnetic $J=-1$ at the $J+K =0$ limit.
The opposite cases with anti-ferromagnetic couplings can be analyzed analogously, and we expect similar physics.

\subsection{Anti-Kitaev limit} \label{subsec:anti-K}
In the limit $J+K = 0$ with $J=-1$, Eq.~\eqref{eq:model} becomes
\begin{align}\label{eq:anti-K_limit}
	H_{\text{anti-K}} =  - \sum_{\langle ij \rangle_\gamma} \sum_{\alpha\neq\gamma} S_i^{\alpha} S_j^{\alpha},
\end{align}
which eliminates couplings between the $S^{\gamma}$ components in the lattice direction $\hat{\mathbf{e}}_\gamma$.
For example, the local Hamiltonian on a $z$-bond reads $H_z = -S_i^x S_j^x - S_i^y S_j^y$.
We refer to this as the anti-Kitaev limit.
The reduced Hamiltonian Eq.~\eqref{eq:anti-K_limit} is invariant upon transforming $S^\gamma \rightarrow -S^\gamma$ for all spins in a spatial $\alpha\beta$-plane orthogonal to $\hat{\mathbf{e}}_\gamma$, as visualized in Fig.~\ref{fig:model}.
The symmetry acts on $d=2$ submanifolds and leads to a subextensive GSD $\propto 2^{L^{D-d}} = 2^L$ for linear system size $L$.

The precise ground states are found by examining the local environment of individual spins.
The local minimal energy at a site $\mathbf{i}$ is minimized ferromagnetically as
$E_\mathbf{i} = -2\big[(S^x)^2 + (S^y)^2 + (S^z)^2\big] = -2|\vec{S}|^2$ (double counting removed).
This means its ground state manifold has an accidental $O(3)$ symmetry. 
Because this $O(3)$ is not a symmetry of $H_{\text{anti-K}}$, it will be lifted by entropy effects via the order-by-disorder mechanism~\cite{Villain80, Henley89}.
The selected states are illustrated in Fig.~\ref{fig:model}: spins form collinear configurations by spontaneously picking up a common axis $c\in\{x,y,z\}$. 
Nevertheless, due to the subsystem symmetry, a long-range spin order is only possible within a spatial plane orthogonal to $\hat{\mathbf{e}}_c$.  

This model provides an example of hybrid symmetry breaking: the system will appear magnetic if observing along directions within an ordered plane but non-magnetic in the perpendicular direction. 
The corresponding order parameters can be constructed as follows,
\begin{flalign}
m_{P,r_c}^{c} &\coloneqq\frac{1}{L^2} \sum_{(r_a, r_b) \in P} S^c_{r_a \hat{\mathbf{e}}_a + r_b\hat{\mathbf{e}}_b + r_c \hat{\mathbf{e}}_c}, \label{eq:mP} \\
Q^{cc} &\coloneqq \frac{1}{L} \sum_{r_c} \left(\, m_{P,r_c}^c m_{P,r_c}^c - \frac{1}{3}\, \right), \label{eq:Q}
\end{flalign}
where $S^c$ denotes the spin component forming the magnetic order, and $a, b, c = x, y , z$ are mutually exclusive.
The two order parameters $m_{P,r_c}^{c}$ and $Q^{cc}$ measure the ordering within and between planes, respectively, and their codimensions ${\rm codim}\left(m_{P,r_c}^{c}\right) = 2$ and ${\rm codim}\left(Q^{cc}\right) = 1$, add up to the global dimension $D=3$.
The quadratic quantity $Q^{cc}$ may be viewed as an analog of the nematic order in a uniaxial liquid crystal~\cite{BookDeGennes}.  
However, its building blocks here are not local director fields but sub-dimensional macroscopic objects $m_{P,r_c}^{c}$.
This also indicates that the associated phase transition will not follow a usual Landau theory of two order parameters.

\subsection{Kitaev limit} \label{subsec:K}
We next discuss the $J=0, K = -1$ limit
\begin{align} \label{eq:K_limit}
H_{\rm K} =  - \sum_{\langle ij \rangle_\gamma} S_i^{\gamma} S_j^{\gamma},
\end{align}
where spins couple solely through the $\gamma$ component in a $\gamma$-bond.
This reduced Hamiltonian $H_{\rm K}$ is also known as a $t_{2g}$ compass model~\cite{Brink04}, and its form is akin to Kitaev's honeycomb model~\cite{Kitaev06}.
Due to the cubic geometry, it features a $d=1$ subsystem symmetry by flipping $S^\gamma$ in an entire $\gamma$-line of the lattice and consequently a subextensive GSD $\propto 2^{L^2}$.

The ground states of $H_{\rm K}$ can be analyzed in a similar way as in the case of $H_{\text{anti-K}}$.
The energy per site is $E_g = -|\vec{S}|^2=-1$ indicating an accidental $O(3)$ symmetry.
Stable ground states are magnetically ordered, decoupled chains as illustrated in Fig.~\ref{fig:model}.
More concretely, spins form line magnetizations (rank-$1$) thanks to a thermal order-by-disorder. 
The $d=1$ subsystem symmetry ensures no magnetic order beyond one chain.
The magnetic chains have to pack into planes towards one of their two orthogonal directions, leading to a planar nematicity (rank-$2$) for the inter-chain ordering.
The inter-plane order further induces a rank-$4$ quantity as the two orthogonal directions are picked randomly in different planes.

Thus, three order parameters are demanded to characterize the states fully,
\begin{flalign}
&m^a_{L,(r_b, r_c)} \coloneqq \frac{1}{L} \sum_{r_a} S_{r_a\hat{\mathbf{e}}_a + r_b\hat{\mathbf{e}}_b + r_c\hat{\mathbf{e}}_c}^a \label{eq:mL} \\
&Q^{aa}_{L, r_c} \coloneqq \frac{1}{L}  \sum_{r_b} \left(m^a_{L,(r_b, r_c)} m^a_{L,(r_b, r_c)} - \frac{1}{3}\right) \label{eq:QL}\\
&Q^{aabb}_4 \coloneqq \frac{1}{L}  \sum_{r_c} \left(Q^{aa}_{L, r_c} - Q^{bb}_{L, r_c} \right)^2, \label{eq:Q4}
\end{flalign}
which all have a codimension-$1$.
The mutually exclusive $a,b,c$ are determined spontaneously in the hybrid symmetry breaking, while $(r_b, r_c)$ and $r_c$ index the selected lines and planes, respectively.
$Q^{aa}_{L}$ is also a nematic-like order but now made from chain magnetizations $m^a_L$.
Correspondingly, $Q^{aabb}_4$ is analogous to a generalized biaxial nematic order~\cite{Nissinen16}.

The hybrid symmetry-breaking patterns are richer than in the previous example.
The rank-$4$ order $Q^{aabb}_4$ breaks the spin and spatial permutation symmetries with a global plane orientation. 
The four-fold rotation symmetry of each square plane is further broken down to a two-fold one by the direction of $Q^{aa}_{L}$, whereas no nontrivial symmetries are left in a magnetic chain.
We emphasize that these are not three sequential phase transitions but the same phase transition observed at different submanifolds of the system.

\section{First-order finite size scaling}\label{sec:scaling}
First-order phase transitions may be routinely considered less interesting as they do not incur criticality but reflect a sharp change in the system.
However, a recent work Ref.~\cite{Mueller14} showed that first-order phase transitions involving subsystem symmetries deviate from the conventional $L^{-D}$ finite-size scaling.
In this section we briefly review the standard~\cite{Fisher82, Lee91, Janke03} and non-standard~\cite{Mueller14} first-order scalings.
For simplicity we will follow the arguments in Ref.~\cite{Janke03} and consider a periodic boundary condition (PBC), but we refer to Ref.~\cite{Lee91} for a more comprehensive discussion and Ref.~\cite{Borgs95} for the effects of open boundaries.

Assume the system undergoes a single order-disorder transition, where the ordered phase has $q$ equivalent degenerate states.
We denote the thermodynamical free energy density of the ordered and disordered phases as $f_o$ and $f_d$, respectively.
The possibility of the system in a particular ordered or disordered state can be estimated by Boltzmann factors
\begin{flalign}
p_o &\propto e^{-\beta L^D \hat{f}_o}, \\
p_d &\propto e^{-\beta L^D \hat{f}_d}.
\end{flalign}
Then the weights of the two phases are given by $W_o \propto q e^{-\beta L^D \hat{f}_o}$ and $W_d \propto e^{-\beta L^D \hat{f}_d}$, with $W_o+W_d = 1$.

A first-order phase transition occurs at the inverse temperature $\beta$ where $W_o = W_d$.
This fact can be understood from the behaviors of the specific heat $C_V$.
For simplicity, we ignore fluctuations within a phase and denote energy density of the order and disordered phases by $E_o$ and $E_d$, respectively.
The $n$-th moment of the energy density for an arbitrary state can then be approximately estimated by a weighted average~\cite{Janke03}, 
$\langle E^n \rangle \simeq W_o E^n_o + W_d E^n_d$.
The specific heat hence becomes
\begin{flalign}\label{Cv}
C_V &= \beta^2 L^D \left(\langle E^2 \rangle - \langle E \rangle ^2\right) \nonumber \\ 
&= \beta^2 L^D W_o (1 - W_o) (E_d - E_o)^2,
\end{flalign}
which peaks at $W_o=W_d=\frac{1}{2}$.

Thus at a first-order phase transition, one has
\begin{align}\label{eq:PT}
	0 = \ln{(W_o/W_d)} \simeq \ln{q} + L^D \beta (f_d - f_o).
\end{align}
Using $f = E - TS$, where $S$ is the entropy density, and Taylor expanding Eq.~\eqref{eq:PT} around the thermodynamical ($L \rightarrow \infty$) transition point $\beta^{\infty}$, we obtain in leading order    
\begin{align}\label{eq:D_scaling}
\beta_c(L) = \beta_c^{\infty} - \frac{\ln{q}}{L^D(E_d - E_o)} + \mathcal{O}(\frac{1}{L^{D-1}}).
\end{align}
Namely, for a constant GSD, which is the case when breaking a global symmetry, the leading finite-size correction is $\propto L^{-D}$.

The above analyses can be straightforwardly generalized to subsystem symmetry breaking by accordingly taking into account the GSD's size dependence, which is the central idea of the non-standard first-order scaling~\cite{Mueller14}.  
Specifically, for a $d$-dimensional subsystem symmetry, the number of equivalent ground states $q$ grows exponentially with $L^{D-d}$. 
In consequence, Eq.~\eqref{eq:D_scaling} should be modified as 
\begin{align}\label{eq:d_scaling}
\beta_c(L) = \beta_c^{\infty} + \frac{b}{L^{-d} (E_d - E_o)} + \mathcal{O}(\frac{1}{L^{-d-1}}),
\end{align}
where the constant parameter $b$ depends on specific global symmetries and, up to the leading order, affects only the slope of the scaling.

The non-standard scaling Eq.~\eqref{eq:d_scaling} has been verified for the 3D plaquette Ising model and an anisotropically coupled 3D Askin-Teller model~\cite{Mueller14,Johnston17}; both have a $d=2$ plane-flip symmetry.
As the essential idea in the above discussion is the sub-extensive degeneracy, we expect it should generally hold in lattice models with subsystem symmetries.
Nevertheless, the phase transitions can experience stronger finite-size effects due to the slower vanishing of correction terms.

\begin{figure*}[!tb]
\centering
  \includegraphics[width=1.0\textwidth]{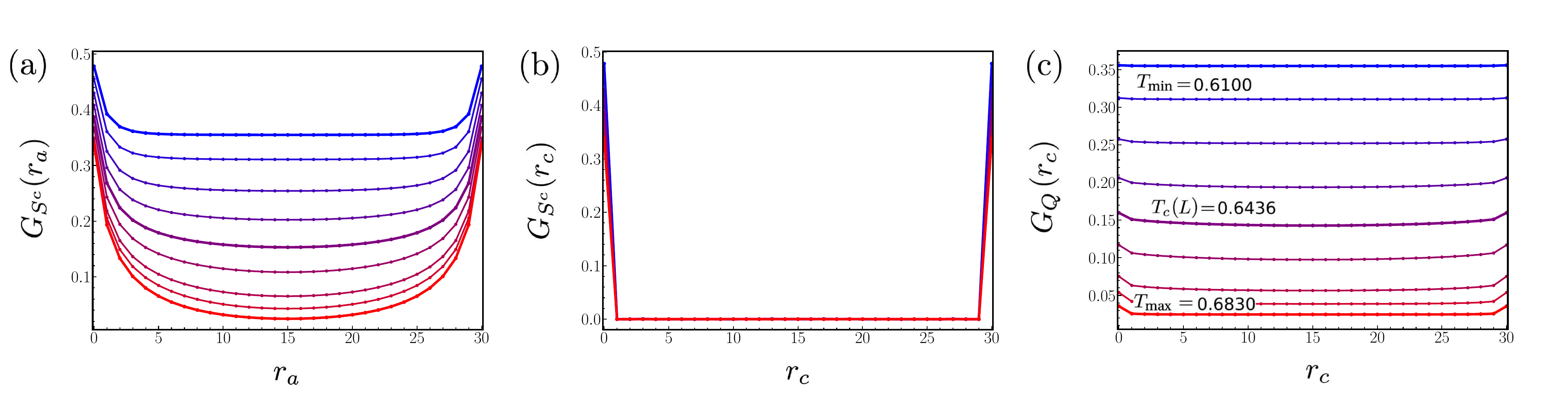}
  \caption{Correlation functions for $H_{\text{anti-K}}$ at various temperatures below (blue) and above (red) the phase transition.
  	(a) Spin-spin correlation along one of the lattice basis vectors in the ordered planes. At low temperatures, a long-range magnetic correlation between the $S^c$ components arises. The symmetric form of the curves comes from periodic boundary conditions.
  	 (b) No magnetic correlation is seen between the planes. 
  	 (c) Nematic correlation in the direction orthogonal to the planes, with a long nematic-like order at low temperatures. An $L=30$ system is taken for example. $N_T = 56$ temperatures are simulated in a range $T \in [0.610, 0.683]$; not all curves are shown for better visibility.}
  \label{fig:AntiKitCorr}
\end{figure*}

\begin{figure}[bt]
  \includegraphics[width=.5\textwidth]{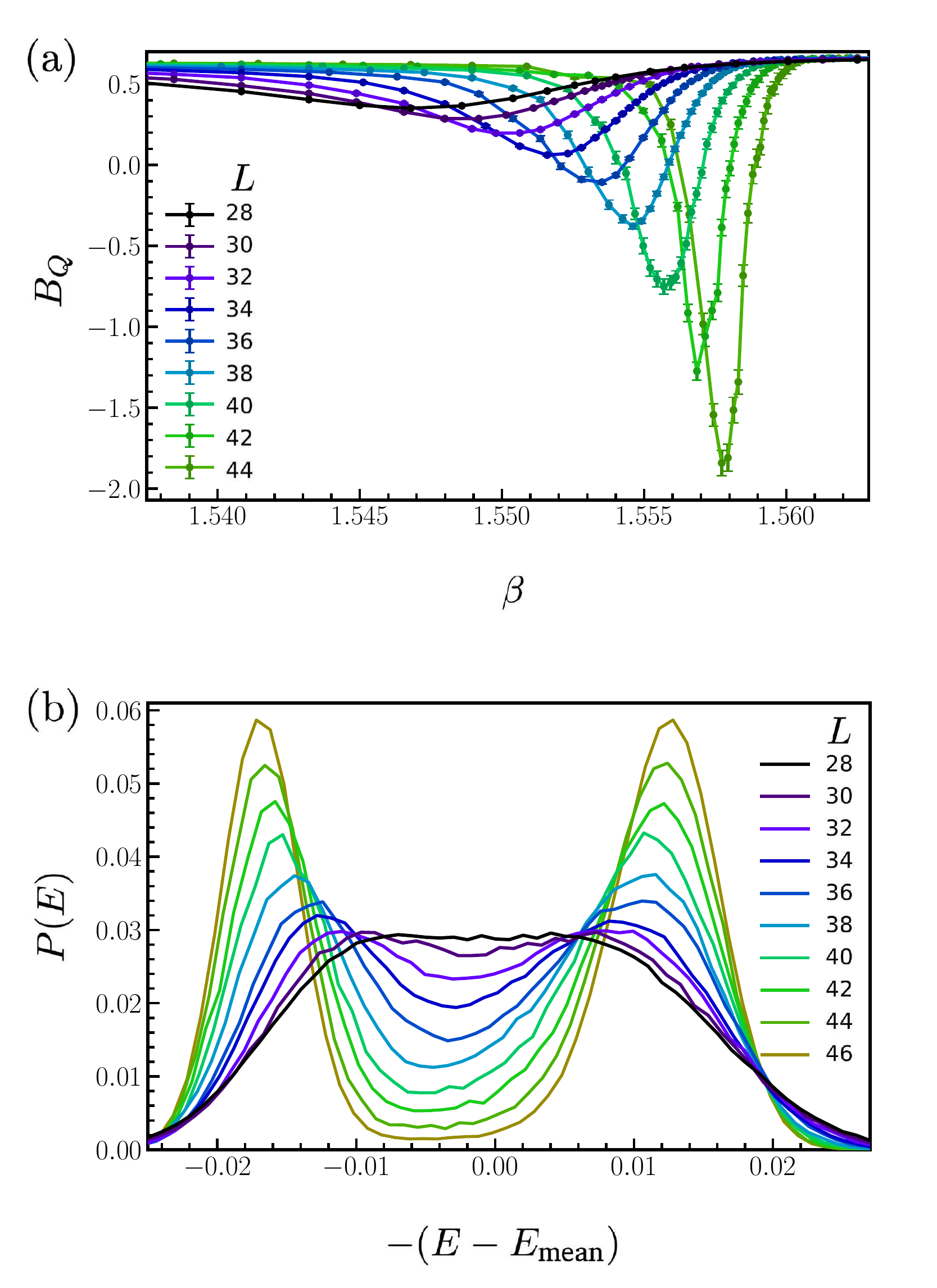}
  \caption{Binder cumulant $B_Q$ (a) and reweighted energy density histogram $P(E)$ (b) near the phase transition of $H_{\text{anti-K}}$. The pronounced Binder dips and double histogram peaks confirm a first-order transition. Equal peak heights are found by attaching an additional weight $e^{- (\beta - \beta^{\text{eqh}}) EV}$ to a raw histogram at inverse temperature $\beta$. $P(E)$ is renormalized and shifted by its mean energy density. The minima between peaks fall slightly away from zero, indicating low- and high-temperature peaks in general have different weights at finite lattice sizes.}
  \label{fig:AntiKitPT}
\end{figure}

\begin{figure}[t]
\centering
  \includegraphics[width=.5\textwidth]{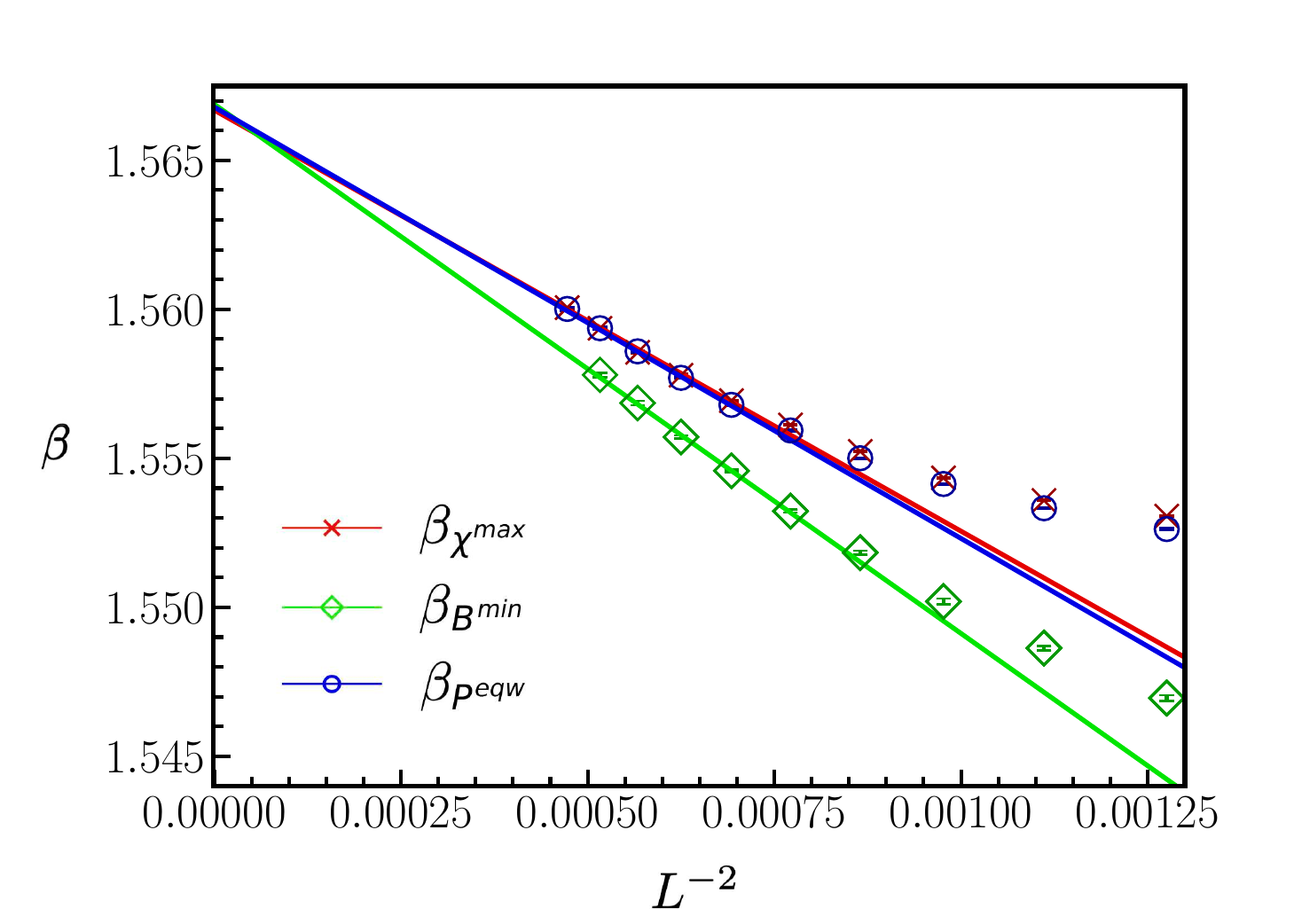}
  \caption{Fits of $H_{\text{anti-K}}$ using the non-standard first-order scaling equation $\beta_c(L) = \beta_c^{\infty} - a/L^{2}$. The finite-size inverse transition temperatures $\beta_c(L)$ are estimated from the extrema of $\chi$ and $B_Q$ and the equal-weight peaks of $P(E)$. The largest five sizes are considered for the fits to reduce finite size effects.}
\label{fig:AntiKitfit}
\end{figure}

\section{Phase transition in the anti-Kitaev limit}\label{sec:anti-Kitaev}
We first discuss the order and phase transition in the anti-Kitaev limit.
We simulate $H_{\text{anti-K}}$ with Monte Carlo techniques by jointly using parallel tempering~\cite{Hukushima96}, heat-bath, and microcanonical overrelaxation updates~\cite{Brown87}.
$N_T = 56$ temperatures are simulated in the proximity of the phase transition, whose distribution is fine tuned to maximize iterations of the replicas~\cite{Katzgraber06}.
Simulations are carried out for linear lattice sizes up to $L=46$ ($\sim 10^5$ spins) under PBC.
We typically run $10^7$ Monte Carlo sweeps and determine the equilibration by comparing results of uniform and random initializations.
The system is considered equilibrated when expectation values of all observables using different initializations agree within error bars.

The hybrid symmetry breaking in $H_{\text{anti-K}}$ and the associated order parameters are confirmed by measuring the disconnected correlators
\begin{flalign}
&G_{S^c} \left(\mathbf{r}_{\in P} \right) = \frac{1}{L^3} \sum_{\{\mathbf{i}\}}  \langle S^c_\mathbf{i} S^c_{\mathbf{i}+\mathbf{r}_{\in P}}\rangle, \\
&G_{S^c}(\mathbf{r}_c) = \frac{1}{L^3} \sum_{\{\mathbf{i}\}}  \langle S^c_\mathbf{i} S^c_{\mathbf{i} + \mathbf{r}_c}\rangle, \\
&G_Q(\mathbf{r}_c) = \frac{1}{L} \sum_{i_c} \langle \left(m_{P,\mathbf{i}}^c\right)^2 \left(m_{P,\mathbf{i} + \mathbf{r}_c}^c\right)^2\rangle.
\end{flalign}
Here the ordered spin component $S^c$ varies over samples but is easily identified by selecting an arbitrary site $\mathbf{i}=(i_x, i_y, i_z)$ and comparing $m_P$ for all three intersecting planes. 
These correlators relate to the order parameters as
$\lim\limits_{|\mathbf{r}_{\in P}| \rightarrow \infty} G_{S^c} \left(\mathbf{r}_{\in P} \right) \rightarrow \langle m_P^c \rangle^2$
and $\lim\limits_{|\mathbf{r}_c| \rightarrow \infty} G_{Q}(\mathbf{r}_c) \rightarrow \langle Q^{cc} \rangle^2$, while $G_{S^c}(\mathbf{r}_c)$ is introduced for reference. 

\begin{table}[tb]
\centering
\caption{Comparison between the standard ($1/L^3$) and non-standard ($1/L^2$) first-order scaling. 
The largest five lattice sizes in Figs.~\ref{fig:AntiKitPT} and ~\ref{fig:AntiKitfit} are considered for fitting.
The validity of the fits is evaluated by the $\chi$-square test, where smaller values of $\chi^2_{\rm dof}$ in general indicate better goodness.
The best fits compatible with a $95\%$ confidence interval are in bold.}
\begin{tabular}{c | c c | c c }
 	\multicolumn{3}{c} {$\qquad\qquad\, 1/L^2$}  & \multicolumn{2}{c} {$1/L^3$}\\
    \arrayrulecolor{white}
    \arrayrulecolor{black}  
    \toprule \hline
    \arrayrulecolor{white}\hline\hline
    \arrayrulecolor{black}  
	$\quad$ Fit $\quad$ 									& $\qquad$ $\beta_{\infty}$ $\qquad$	&  $\,\,\,$ $\rchi^2_{\text{dof=3}}$	$\,\,\,$ & $\qquad$ $\beta_{\infty}$ $\qquad$	& $\,\,\,$ $\rchi^2_{\text{dof=3}}$ $\,\,\,$\\
    \hline
	$\beta_{C_V^{\text{max}}}$ 					& 1.566532(33) 		& 5.28			& 1.563882(15) 		& 11.03			\\ 
	$\beta_{\rchi^{\text{max}}}$ 				& 1.566685(21) 		& 6.30			& 1.563987(10)		& 13.84			\\
	$\beta_{B^{\text{min}}}$  	& $\mathbf{1.566883(91)}$ 		& $\mathbf{1.60}$			& 1.563073(34)		& 5.18			\\
	$\beta_{H^{\text{eqh}}}$ 			& 1.566609(31) 		& 5.52			& 1.563968(15)		& 11.29			\\ 
	$\beta_{H^{\text{eqw}}}$ 			& $\mathbf{1.566787(61)}$ 		& $\mathbf{2.29}$			& 1.563927(25)		& 6.38			\\
    \arrayrulecolor{white}\hline\hline
    \arrayrulecolor{black}  
    \hline \bottomrule
\end{tabular}
\label{tab:AntiKit}
\end{table}

\begin{figure*}[!t]
\centering
  \includegraphics[width=1\textwidth]{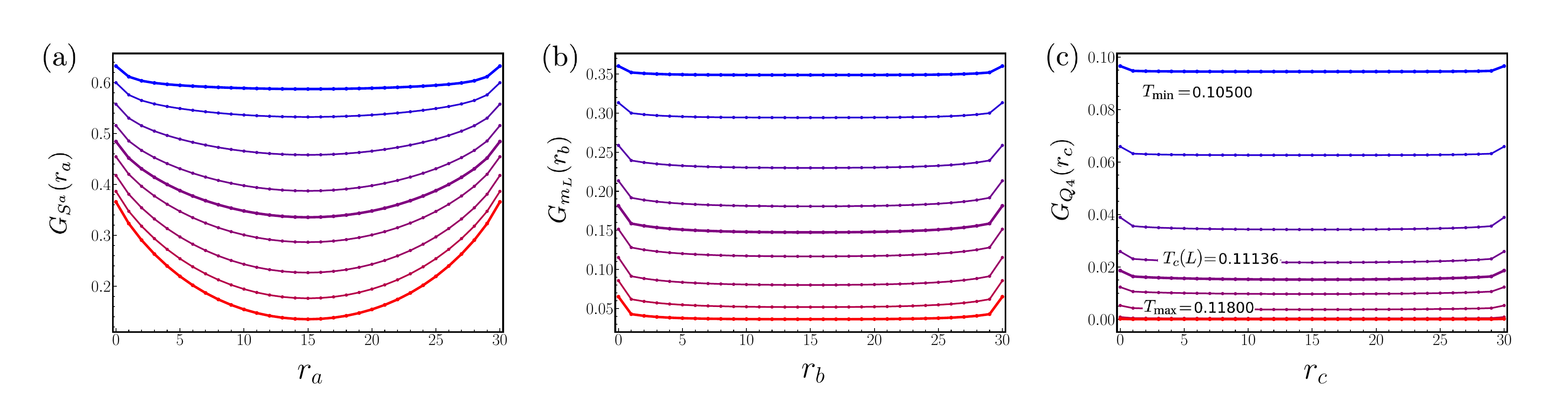}
  \caption{Correlation functions in the Kitaev limit for $L=30$ system below (blue) and above (red) the phase transition. The intra-chain spin correlation $G_{S}$, inter-line nematic correlation $G_{m_L}$, and inter-plane rank-$4$ correlation $G_{Q_4}$ are measured along their respective ordering directions. Each of them shows a long range order at low temperatures. The simulated temperature interval is indicated by $T_{\rm max}$ and $T_{\rm min}$.}
 \label{fig:KitCorr}
\end{figure*}

We measure the three correlators for several temperatures around $T_c$ at a lattice $L=30$. 
As shown in Fig.~\ref{fig:AntiKitCorr}, the spin correlators $G_{S^c}$ only exhibit a long-range correlation along directions $\mathbf{r}_{\in P}$.
In the perpendicular direction $\mathbf{r}_c \perp P$, the spin-spin correlation vanishes but the planar nematic correlator $G_{Q}$ is long-ranged.
This confirms the hybrid symmetry breaking, as one will observe distinct symmetry breaking patterns when looking at different submanifolds of the system.

To further understand the associated phase transition, we examine behaviors of the energy histogram $P(E)$, specific heat $C_V$, susceptibility $\chi$, and Binder cumulants $B$, where
\begin{flalign}
&P(E) = \langle\delta(E-E^\prime)\rangle, \\
&\chi_{\mathcal O} = \frac{L^D}{T}\left(\langle\mathcal O^{\,2} \rangle - \langle \mathcal O \rangle^2\right), \\
&B_{\mathcal O} = 1- \frac{\langle \mathcal O^{\,4}\rangle}{3\langle \mathcal O^{\,2}\rangle^2}.
\end{flalign}
and $\mathcal{O} = m_P^c, \, Q^{cc}$.
Since there is only a single phase transition, both order parameters can detect $T_c$ and its scaling.
Nonetheless, we observe a slightly weaker finite-size effect in $Q^{cc}$ and hence use it as the major order parameter.
It may also be useful to mention that if $\mathcal{O}$ contains a finite trace, namely, if the factor $\frac{1}{3}L$ was not subtracted in Eq.~\eqref{eq:Q}, the Binder cumulant would approach $\frac{2}{3}$ in both the low- and high-temperature limits, instead of vanishing as $T \rightarrow \infty$. 

These quantities reliably determine the order of a phase transition.
In particular, near a first-order transition, the histogram $P(E)$ will show two peaks representing the weights, $W_o$ and $W_d$, of the ordered and disordered phases~\cite{Challa86, Lee90}.
Correspondingly, the curve of $B_{\mathcal O}(T)$ non-monotonically depends on temperatures with a minimum dipping at an effective transition point~\cite{Binder84}.
Such behaviors are observed for $H_{\text{anti-K}}$, as we plot in Fig.~\ref{fig:AntiKitPT}. 
The double $P(E)$ peaks and Binder dip become evident from $L \sim 30$ and evolve towards $\delta$ functions when increasing system sizes, which unambiguously confirms the first-order nature of the transition.   

Effective transition temperatures $T_c(L)$ at a given $L$ can be estimated from locations of the $C_V$ and $\chi$ peaks and that of the Binder dip.  
The energy histogram provides two additional estimators: $T_c^{\rm eqw}(L)$ and $T_c^{\rm eqh}(L)$, where $P(E)$'s two peaks have the same weight (${\rm eqw}$) and height (${\rm eqh}$), respectively.
The equal-weight criterion, $W_o = W_d$, is understood by its relation to the $C_V$ maximum as discussed in Sec.~\ref{sec:scaling} and can be found by minimizing
\begin{align}
\left[\Delta W(T)\right]^2  &= \left[ W_o(T) - W_d(T) \right]^2 \nonumber \\
& = \left[\sum_{E < E_{\text{min}}} P(E,T) - \sum_{E > E_{\text{min}}} P(E, T)\right]^2,
\end{align}
with $E_{\text{min}}$ being the valley minimum between the two peaks.
The estimator of $T_c^{\rm eqh}(L)$ is more empirical; its common use in the literature may be because an equal-height behavior can be easily found through reweighting methods.
These effective temperatures typically suffer different degrees of finite-size effects but shall converge when approaching $L \rightarrow \infty$.

To determine the infinite volume $T_c^\infty$,  or equivalently, its inverse $\beta_c^\infty$ and the leading-order finite-size correction, we fit $\beta_c(L)$ obtained from different estimators against single-term powers $L^{-2}$ and $L^{-3}$.
The five largest sizes ($L \geq 36$) are used to ensure $L$ sufficiently exceed the correlation length. 
As summarized in Table~\ref{tab:AntiKit}, we observe that the $L^{-2}$ fitting leads to noticeably more consistent values of $\beta_c^\infty$, consolidating the non-standard scaling relation Eq.~\eqref{eq:d_scaling}. 
Fits with the best goodness are plotted in Fig.~\ref{fig:AntiKitfit}, and we extrapolate $\beta_c^\infty \approx 1.5667(2)$.
We expect including $L^{-3}$ as a sub-leading term can further improve the results, while such a two-term fitting also requires more data points.

It is interesting to know whether the accidental symmetry and its entropic lifting play a critical role in the nature of the phase transition.
For this purpose we replace the $O(3)$ spins in Eq.~\eqref{eq:anti-K_limit} with discrete spin variables 
$\widetilde{S}_i = (\pm 1, 0, 0),  (0, \pm 1, 0),  (0, 0, \pm 1)$ that are transformed by the point group $C_{3i}$.
By doing so, accidental symmetries and soft modes become irrelevant, and the resulting Hamiltonian $\widetilde{H}_{\text{anti-K}}$ formally has the same symmetries and order parameters as $H_{\text{anti-K}}$.
It turns out $\widetilde{H}_{\text{anti-K}}$ shows even stronger jumps in energy and order parameters.
This may imply that the observed first-order nature of the phase transitions in the two Hamiltonians may relate to the presence of the planar subsystem symmetry.

\section{Phase transition in the Kitaev limit}\label{sec:Kitaev}
We now discuss the hybrid symmetry breaking in the $H_{\rm K}$ limit.
This Hamiltonian was also studied in Ref.~\cite{Gerlach15} based on energy related quantities and mostly under a screw PBC that explicitly breaks line-flip symmetries and modifies the ground-state manifold.
In this section, we consider a regular PBC preserving the subsystem symmetry and characterize its phase transition from the viewpoints of order parameters.
Simulations are performed utilizing the same Monte Carlo methods as in Sec.~\ref{sec:anti-Kitaev}, with system sizes up to $L=40$. 

As analyzed in Sec.~\ref{subsec:K}, the robust ground states of $H_{\rm K}$ comprise decoupled magnetic chains.
Normally, a 1D long-range order is prohibited for classical spins at finite temperature due to the Mermin-Wagner theorem~\cite{Mermin66}.
The chain magnetization here is nevertheless possible thanks to order-by-disorder and entropic lifting of the accidental $O(3)$ degeneracy.
A similar entropy driven 1D magnetization was also reported in a 2D classical compass model~\cite{Mishra04,Wenzel08}.
For a consistency check, we also verified with separate Monte Carlo simulations the absence of a phase transition when replacing the $O(3)$ spins with discrete $C_{3i}$ spins.

The in-chain magnetization $m_L$, line nematicity $Q_L$, and inter plane higher-rank nematic order $Q_4$ defined in Eqs.~\eqref{eq:mL}--~\eqref{eq:Q4} lead to a rich set of (high-rank) disconnected correlators:
\begin{flalign}
&G_{S^a} (r_a) = \frac{1}{L^3}  \sum_{\mathbf{i}} \langle S^a_{\mathbf{i}} {S^{a}}_{\mathbf{i} + r_a \hat{\mathbf{e}}_a}\rangle, \\
&G_{m_L} (r_b) = \frac{1}{L^2}  \sum_{(i_b,i_c)} \langle (m^a_{L,(i_b,i_c)})^2 (m^a_{L,(i_b,i_c) + r_b \hat{\mathbf{e}}_b})^2  \rangle \\
&G_{Q_4}(r_c) = \frac{1}{L}  \sum_{i_c}  \langle (Q^{aa}_{L, i_c} - Q^{bb}_{L, i_c})^2 (Q^{aa}_{L, i_c + r_c} - Q^{bb}_{L, i_c + r_c})^2 \rangle,
\end{flalign}
along their respective ordering directions.
To probe these directions in a given sample, one can arbitrarily choose a site $\mathbf{i} = (i_x, i_y, i_z)$ and compute the three possible line magnetizations passing through it.
The one with the largest magnitude determines the relevant $m^a_{L,(i_b,i_c)}$ for this site. 
One next compares the potential $Q^{aa}_{L, i_c}$ along the two remaining directions $\hat{\mathbf{e}}_b$ and $\hat{\mathbf{e}}_c$, whose result then fixes the common orientation of ordered planes.
Nevertheless, since planes are decoupled in ground states, the line magnetizations in different planes remain to be determined individually (see Fig.~\ref{fig:model}).

\begin{figure}[!t]
\centering
  \includegraphics[width=0.5\textwidth]{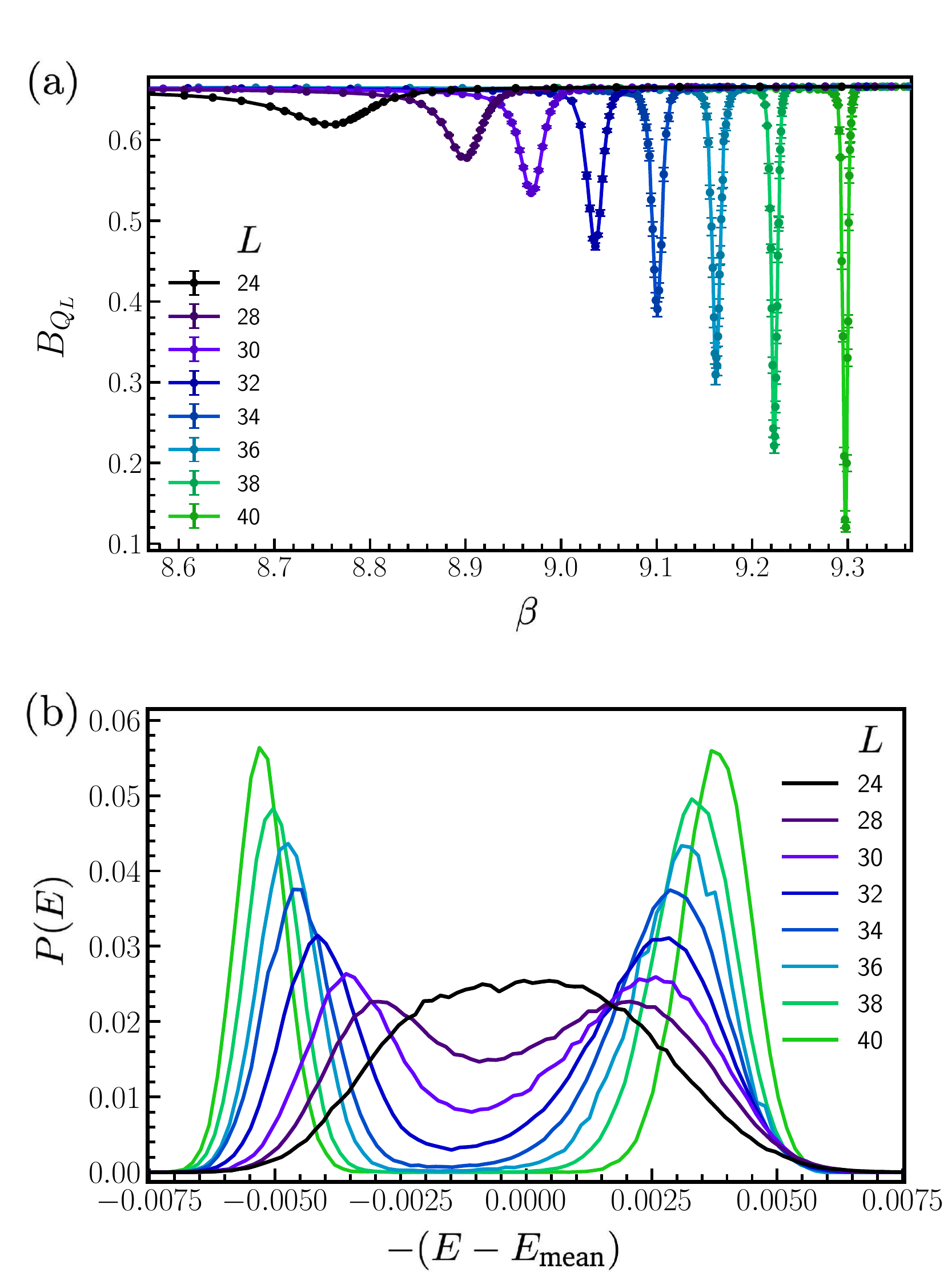}
  \caption{Binder cumulant $B_{Q_L}$ (a) and reweighted energy density histogram $P(E)$ (b), detecting a first-order phase transition in $H_{\rm K}$. The in-line magnetization $m_L$ and rank-$4$ order $Q_4$ give consistent results.}
   \label{fig:KitPT}
\end{figure}

\begin{figure*}[!t]
\centering
  \includegraphics[width=1\textwidth]{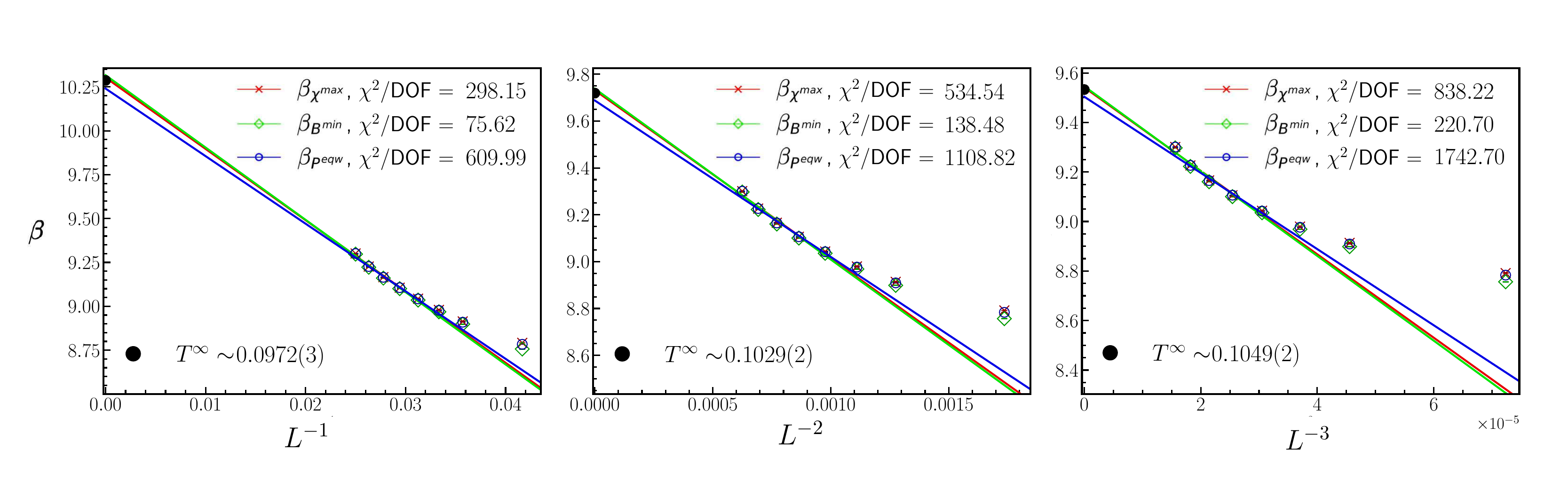}
  \caption{Comparison of the fits at the Kitaev limit using the scaling equation $\beta_c(L) = \beta_c^{\infty} - a L^{-d^\prime}$, with $d^\prime = 1,2,3$ and $24 \leq L \leq 40$. The inverse finite-size transition temperatures $\beta_c(L)$ are estimated from the extrema of $\chi_{Q_4}$ and $B_{Q_L}$ and the equal-weight peaks of $P(E)$. The largest five sizes are considered for the fitting. Single-term scaling cannot be established with accessible system sizes. Nevertheless, the extrapolations confirm a finite-temperature phase transition  around $T^{\infty} \sim 0.1$ and justify the deviation to the standard $L^{-3}$ scaling.}
   \label{fig:Kitfits}
\end{figure*}

In Fig.~\ref{fig:KitCorr} we measure the three correlators around the phase transition for a lattice $L=30$.
At temperatures below $T_c$, they all lead to long-range order.
The faster convergence of $G_{Q}$ and $G_{m_L}$ with distance can be understood from their higher ranks.
We also verified no long-range correlations if the correlators were measured in a wrong direction, as in Fig.~\ref{fig:AntiKitCorr} for the anti-Kitaev limit.
This hence confirms three distinct symmetry-breaking patterns in different submanifolds of the system.

The nature of the phase transition is again detected by the energy density histogram $P(E)$ and Binder cumulant $B_{\mathcal O}$.
As shown in Fig.~\ref{fig:KitPT}, $H_{\rm K}$ also experiences a first-order transition.

We further crosscheck the behaviors of $C_V$, $P(E)$, $\chi_{\mathcal O}$, and $B_{\mathcal O}$, with $\mathcal{O} = m_L^a, \, Q_L^{aa}, \, Q_4^{aabb}$, to find the thermodynamical transition temperature and its scaling. 
According to the non-standard scaling relation Eq.~\eqref{eq:d_scaling}, the leading contribution here should be given by $L^{-1}$ instead of $L^{-3}$.
Nevertheless, we observe very strong finite-size effects, and with system sizes $L\leq 40$, none of the single-term scaling $L^{-d^\prime}$ with $d^\prime =1,2,3$ can give satisfying goodness of fits, as shown in Fig.~\ref{fig:Kitfits}.
A significantly improved fit is achieved when considering a two term scaling $\beta_c(L) = \beta_c^{\infty} + a L^{-1}  + b L^{-2}$, whereas,  given the limited data points, this could also be an artifact of overfitting.
Combining the standard $L^{-3}$ scaling under a screw PBC reported in Ref.~\cite{Gerlach15}, a most likely scenario is that the line-flip subsystem symmetry indeed has deflected $H_{\rm K}$ from the standard first-order scaling relation, and $L^{-1}$ may still be the leading order correction.
Nevertheless, a sub-leading $L^{-2}$ term is suppressed very slowly, which makes the single-term scaling not a good approximation for accessible system sizes.

\section{Summary and outlook}\label{sec:conclusion}
Subsystem symmetries provide new possibilities to explore novel states of matter beyond existing frameworks of global and gauge symmetries.
In this work we showed that breaking them can lead to unconventional phases where the system breaks various symmetries in different submanifolds.
We referred to this phenomenon as hybrid symmetry breaking and demonstrated it with the compass-model limits of a classical cubic Heisenberg-Kitaev Hamiltonian (Section~\ref{sec:model_op}).
Because of a planar subsystem symmetry, the low-temperature phase in the anti-Kitaev limit of this model is magnetic within a plane but nematic-like between planes.
Similarly, its Kitaev limit simultaneously possesses an in-chain magnetic order, an inter-chain nematic order, and an inter-plane higher-rank nematic order due to a line-flip symmetry.

We further investigated their phase transitions comprehensively utilizing large-scale Monte Carlo simulations. 
Our results consolidated the non-standard first-order scaling that takes subextensive degeneracies into account.
We found strong evidence that the leading finite-size correction for the anti-Kitaev limit follows an $L^{-2}$ scaling instead of a conventional $L^{-3}$ behavior (Sec. ~\ref{sec:anti-Kitaev}).
We also observed a strong violation of the conventional scaling in the Kitaev limit and argued that single-term fits are not favored due to a prolonged sub-leading correction (Sec.~\ref{sec:Kitaev}).

Our work enriches the scenarios of conventional spontaneous symmetry breaking and sheds light on the influences of subsystem symmetries in phase transitions.
The results are also useful for distinguishing the genuine effects of fractons, which require additional geometrical constraints besides the presence of a subsystem symmetry~\cite{Vijay16}.

Several interesting issues deserve future studies.
The first question is if this hybrid symmetry breaking is relevant for quantum systems.
This is promising, as its occurrence in classical models essentially originates from symmetries.
Another exciting direction is investigating its impact on dynamical and transport behaviors of a system.
It was recently shown that a U(1) subsystem symmetry can lead to anomalous subdiffusion and new hydrodynamical universality classes~\cite{Iaconis19, Gromov20}.
Those subdimensional orders arising from a hybrid symmetry breaking naturally support different quasiparticles and can potentially have distinct diffusion properties.
Moreover, it is also important to understand if there is a generic symmetry reason why known 3D subsystem symmetrical models commonly undergo a first-order phase transition.
Examples aside from the two cases studied here include the two dual spin models of the X-cube code~\cite{Song22,Mueller14}, dual models of the checkerboard and Haah's codes~\cite{Vijay16, Haah11, note_claim}, and a breathing pyrochlore magnet~\cite{Yan20}.
The answer to this question can be crucial for the future development of field theories of subsystem symmetry breaking.

The data and simulation parameters used in this work are available in Ref.~\cite{Data_repo2023}.

\begin{acknowledgments}
G.C., L.P., and K.L. acknowledge support from FP7/ERC Consolidator Grant QSIMCORR, No. 771891.
This project is partially funded by the Deutsche Forschungsgemeinschaft (DFG, German Research Foundation) under Germany's Excellence Strategy -- EXC-2111 -- 390814868.
The research is also part of the Munich Quantum Valley, which is supported by the Bavarian Government with funds from the Hightech Agenda Bayern Plus.
Our numerical simulations make use of the TKSVM library~\cite{Greitemann19, Liu19} and the ALPSCore library~\cite{Gaenko17}.
The simulations were performed on the KCS cluster at Leibniz-Rechenzentrum (LRZ) and the ASC cluster at Arnold Sommerfeld Center.
\end{acknowledgments}

\bibliographystyle{apsrev4-1}
\bibliography{subsystem.bib}
\end{document}